\documentclass[sigconf]{acmart}

\AtBeginDocument{%
  }

\copyrightyear{2024}
\acmYear{2024}
\setcopyright{rightsretained}
\acmConference[NordiCHI Adjunct 2024]{Adjunct Proceedings of the 2024 Nordic Conference on Human-Computer Interaction}{October 13--16, 2024}{Uppsala, Sweden}
\acmBooktitle{Adjunct Proceedings of the 2024 Nordic Conference on Human-Computer Interaction (NordiCHI Adjunct 2024), October 13--16, 2024, Uppsala, Sweden}\acmDOI{10.1145/3677045.3685432}
\acmISBN{979-8-4007-0965-4/24/10}

\begin{document}

\title{Co-badge: An Activity for Collaborative Engagement with Data Visualization Design Concepts}

\author{Damla \c{C}ay}
\email{damla.cay@mome.hu}
\affiliation{%
  \institution{Data Storytelling Hub, Moholy-Nagy University of Art and Design}
  \city{Budapest}
  \country{Hungary}
}

\author{Mary Karyda}
\affiliation{%
  \institution{Data Storytelling Hub, Moholy-Nagy University of Art and Design}
  \city{Budapest}
  \country{Hungary}
}

\author{Kitti Butter}
\affiliation{%
  \institution{Data Storytelling Hub, Moholy-Nagy University of Art and Design}
  \city{Budapest}
  \country{Hungary}
}

\renewcommand{\shortauthors}{\c{C}ay et al.}

\begin{abstract}
  As data visualization gains popularity and projects become more interdisciplinary, there is a growing need for methods that foster creative collaboration and inform diverse audiences about data visualization. In this paper, we introduce Co-Badge, a 90-minute design activity where participants collaboratively construct visualizations by ideating and prioritizing relevant data types, mapping them to visual variables, and constructing data badges with stationery materials. We conducted three workshops in diverse settings with participants of different backgrounds. Our findings indicate that Co-badge facilitates a playful and engaging way to gain awareness about data visualization design principles without formal training while navigating the challenges of collaboration. Our work contributes to the field of data visualization education for diverse actors. We believe Co-Badge can serve as an engaging activity that introduces basic concepts of data visualization and collaboration.
\end{abstract}



\begin{teaserfigure}
  \includegraphics[width=\textwidth]{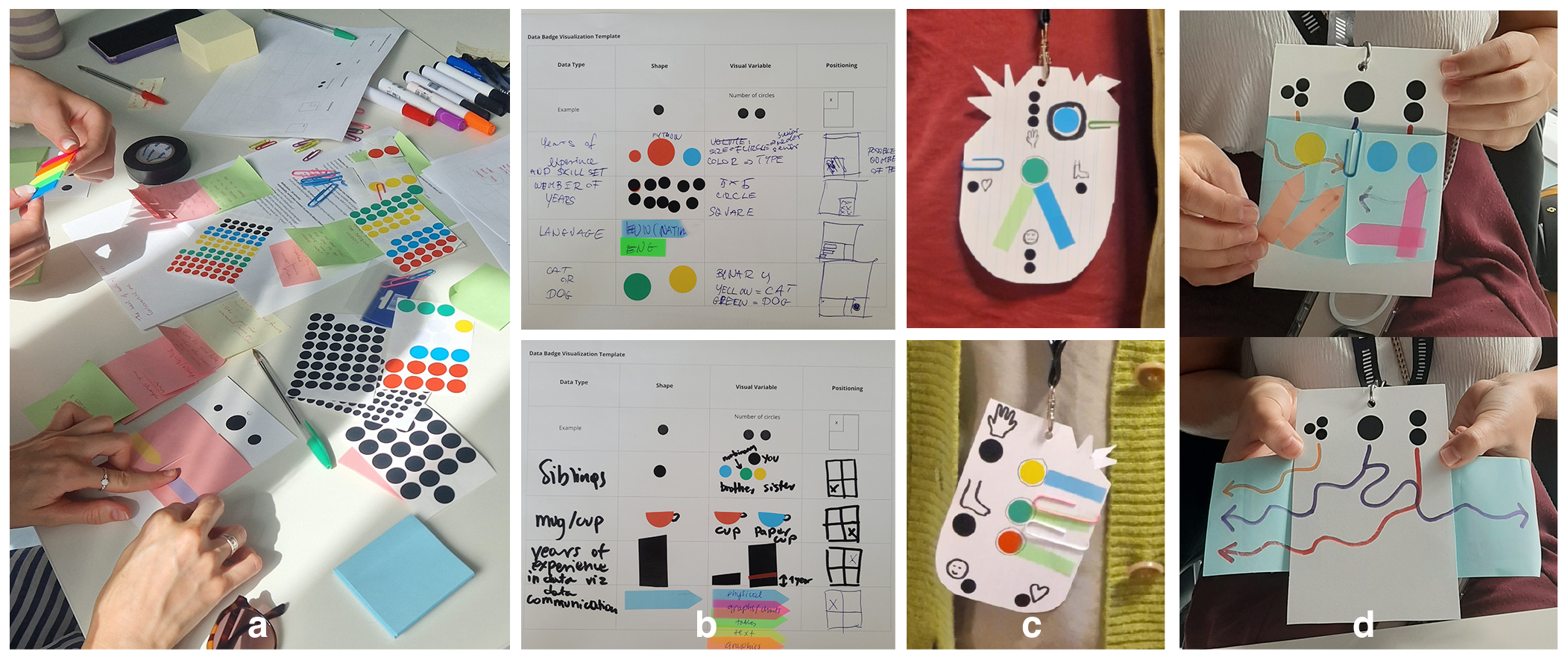}
  \caption{(a) Participants creating their data badges (b) Two examples of the visualization template. (c) Two badges depicting the same information about different participants (d) A badge that has two layers of information about the participant.}
  \Description{Different images showing stages of the workshop, the first image shows participants during the workshop, the second image shows two versions of the visualization template, and the rest of the images show different co-badge examples from the participants.}
  \label{fig:teaser}
\end{teaserfigure}

\received{1 June 2024}
\received[revised]{1 July 2024}
\received[accepted]{20 August 2024}

\maketitle

\section{Introduction and Related Work}

The widespread adoption of data visualization across various fields necessitates collaborative strategies that cater to both experts and novices \cite{ccay2020colvis}. Recognizing the importance of user-centred design, Koh et al., \cite{koh2011developing} emphasize activities that familiarize participants with diverse visualization techniques, marking a shift towards more inclusive collaboration paradigms. As collaborative environments become increasingly interdisciplinary, understanding how to effectively work together in data-driven settings becomes crucial. Bach et al., \cite{bach2023challenges} highlight the necessity of blending core visualization skills with broader competencies like critical thinking, creativity, and collaboration. Echoing this, Kerzner et al. \cite{kerzner2018framework} propose guidelines for workshops that adapt to users' needs, advocating for a visualization mindset. 

The concept of constructive visualizations \cite{huron2014constructive} marks a change in how we create visual representations that are dynamic and flexible. By leveraging the simple task of stacking physical tokens, this approach democratizes processes of visualization creation, making it accessible for non-experts. The simplicity, expressiveness, and dynamic nature of constructive visualization allow for the creation of novel visualizations through processes reminiscent of play, thereby fostering an environment where creativity and exploration take precedence over technical expertise. Data badges \cite{panagiotidou2020data}, a form of constructive visualization, have been explored where participants engage in the individual construction of physicalizations based on predefined data dimensions. This idea was implemented during an academic seminar where participants constructed their professional profiles by arranging physical tokens on a wearable canvas. This hands-on method promoted a group activity that emphasized the creative and social benefits of making data tangible. 

We build upon these works by incorporating collaborative elements into the construction of data badges and allowing people of different personal and professional profiles to define which visual variables fit into their group's perceived identity. In that way, personal data construction is linked to collective decision-making, creating a feeling of ownership but also a shared identity. Thus, we conducted a series of workshops where participants collaboratively engaged in the data visualization process within a creative setting. Co-Badge is a 90-minute design activity where participants collaboratively construct visualizations. By allowing participants to ideate and prioritize relevant data types, map them to visual variables, and collaboratively construct data badges, Co-Badge introduces a playful approach where collaboration, creativity, and, critical thinking are at the centre of data visualization construction. This collaborative activity can serve as an ice-breaker and an educational tool that introduces collaborative data visualization to a diverse audience.

\section{The Co-Badge Workshops}

We conducted the workshop in three different settings: at a summer school, during a university onboarding day, and as part of a team-building event. Our goal in organizing the workshops in different setups was to have individuals with varying backgrounds and experience levels in data visualization, thereby offering diverse insights into the Co-Badge activity. A total number of 74 people participated in the Co-Badge activity. Every workshop session lasted approximately 90 minutes and consisted of six phases: introduction, topic ideation, topic selection, visualization design, crafting of the badge, and reflection. Below we outline the three diverse workshop settings. 

\subsection{Context and Participants}

The first workshop was conducted as part of a Data Storytelling Summer School for professionals organized by Moholy-Nagy University of Art and Design. The 35 participants of this workshop were designers, researchers, data scientists and design students interested in the topic of data visualisation. Most of the participants were already familiar with data visualization principles and were interested in learning how to communicate data in engaging ways. In this workshop, we introduced the Co-Badge activity as an icebreaker easing the participants into working with data but also, as a tool to strengthen the relationships among the people of each group. The second workshop was conducted during a University Onboarding Day. The participants of this session consisted of 16 freshmen from various design disciplines such as design management, media, textile and graphic design. Our participants did not have any previous experience in designing data representations however, they were experienced in creative thinking, co-design, iterative processes and prototyping. The third workshop was part of a team-building event where 23 professionals of different backgrounds - design researchers, administrators, etc – participated. Not all participants had previous experience of creating data representation but they were familiar with creative exploration. Here the workshop was implemented as a creative team-building activity.

\subsection{Workshop Process}
At the start of each workshop, we asked participants to form groups of 4 to 6 people, and each group was assigned to a table. Every table had stationery materials that the participants could use during the activity. This included sticky notes, dot stickers with different sizes and colours, coloured tapes, markers, scissors, and paper clips. After group forming, we introduced the participants to the outline of the Co-Badge activity. We explained the purpose and the schedule of the workshop. We emphasized promoting creative collaboration in a data-driven context without the pressure of arriving at a polished output. 

During ideation, participants individually brainstormed questions and topics that they wanted to know about each other or to share about themselves. They wrote every idea on separate sticky notes. In the selection phase, participants shared the contents of their sticky notes with each other and discussed the possible topics. The sticky notes were then grouped into themes, and themes were named. Then every participant voted on three of their favorite topics using dot stickers. Based on the results of the dot voting, every group decided which topic or topics they wanted to move forward with. By the end of this phase, they arrived at  3 to 6 questions that they planned on visualizing on their badges. After deciding on the questions, they defined what data types they needed to visualize to answer those questions. To help them design the visualizations, we provided every participant with a template \cite{osfCoBadge} that helped them understand the basic principles of linking visual variables to the selected data types. 

The template included a table to fill in with the column titles data type, shape, visual variable and position. We added a filled row as an example. The back side offered additional examples of visual variables (length, size, amount) linked to questions such as \textit{"What type of skills do you have?"}. We provided the participants with brief guidance on aligning visual and conceptual hierarchies and encouraged them to be creative and explore unconventional ways of visualizing data. To create their visual mapping, they filled in a final template that served as a guide for crafting their individual badges. After they finalized the template, everyone had the opportunity to make their badge based on it. They used the stationery materials for crafting and an empty badge that was provided by the facilitators. Upon completion, every group shared the highlights of their process with the rest, as well as the final visualization template, and the badges that they crafted. After this, a post-activity open discussion took place where participants had the chance to comment on the works and the process of the workshop. The main topics discussed were personal experiences, aspects of collaboration, and learning about visualizing data. As part of the reflections, participants answered questions about their workshop experience focusing on aspects like ease, playfulness, materials, process, time, and the provided templates. Suggestions for activity improvement were also encouraged.

For the analysis, our data were notes from observations, photos of the participants’ process and their individual co-badges, audio recordings of post-activity discussions and participants' sticky notes. We conducted open coding to analyse the data. We share all templates, structure and badge examples from participants at the Co-Badge Repository \cite{osfCoBadge}. 

\section{Preliminary Findings}

\subsection{Insights about Data Visualization}

The workshops presented an opportunity for participants to creatively engage with data, highlighting the balance between aesthetics and functionality in data visualization. The playful and experimental approach we followed was appreciated, as P46 said they were \textit{"amazed that it’s so allowing and playful, and it doesn't need to be beautiful."} This sentiment underscores the workshops' role in fostering an environment where the pressure of creating aesthetically perfect outcomes was removed, encouraging exploration and creativity.

In several cases, participants spoke about the principles and challenges of data visualization they realized during the process. For instance, P50 who had experience with urban planning data, noted the challenge of visualizing data for a general audience noting "\textit{This makes you reflect in ways to visualize that information in an easier way."} On a similar note, P70 said
\textit{"It was always important that we are doing something that is easy to understand for people who are looking at it."} These realizations highlight that it is crucial to consider the audience when creating data visualizations. In another case linked to data visualization principles, participants made remarks about the complexity of balancing creative expression with the need for clarity in the visual message. Related to this, P62 stated, "\textit{overcomplicating the visualization loses people's focus}". P22 addressed the challenge of deciding on the visuals saying that \textit{"there are a few nuances, like how should we distribute different visuals so it will look nice given the actual data."}

The workshops surfaced the complexity of translating questions into data and then into visual forms. P58 noted,\textit{ "You've also stimulated us to think a bit more creatively about how to take qualitative questions and present it in a more quantitative way. I think it is also a great challenge"}. This observation was echoed by another participant who noted  \textit{"The way we understood the questions themselves influenced heavily what kind of representation we were offering and that was hard to match. So the visual part was not hard, but to get on the same page of how do we understand that question, how do we understand the quantifiable aspect of that question."} 

\subsection{Collaboration}

The collaborative nature of the workshops fostered a rich learning environment, enabling participants to share ideas and navigate design challenges together. P37 and P12 highlighted the enjoyment and efficiency of collaborating with individuals from diverse backgrounds, leading to a quick consensus on visualization choices. P65 noted, \textit{"Everyone kindly accepted each other's ideas and inputs. So we came very fast to agree on what we wanted to choose. How do we want to visualize it?"}. However, collaboration also presented challenges, especially among participants with similar backgrounds. P41 from Workshop 2 expressed a preference for more diverse group compositions to avoid getting lost in \textit{"unnecessary, jittery parts of the process."} This sentiment was mirrored by P4, who noted the stubbornness within their group, reflecting the complexities of teamwork and the need for compromise. This was observed in a group where all participants were from the same disciplinary background carrying strong opinions about graphical visual aesthetics. In that case, every person in the group created their own data badge failing to present a collective visual identity. Despite that, the competitive yet collaborative atmosphere spurred motivation and inspiration, as participants observed and learned from the creativity and perspectives of others. While our focus was on collaboration within groups, interactions between groups were particularly enriching. Competing and comparing results with other groups not only motivated participants but also exposed them to different perspectives, underscoring the importance of presenting results across teams. 

\subsection{Workshop Process and Materials}
The constraints on time and materials acted as a double-edged sword; while challenging, they also spurred creativity. For instance, P37 found themselves more creative with limited materials, noting \textit{"If you have too much choice, then you are sometimes getting lost in these choices. And if we have only this limited amount, then we become more creative with what we have.}", while P41 who was a graphic design student said that it would have been nice for them if they had different materials or textures to play with, in order to have more visually appealing results. We observed that dot stickers of different shapes and sizes were commonly used. P46 said \textit{didn't have lots of different types of materials that we could use. But this is the thing that made us be a little bit more creative with all these stickers that we had."}. While some participants opted for creating visualizations such as scatterplots using only markers and dot stickers, others even modified the shape of the badge to convey information. Materials such as sticky notes were also used in unexpected and creative ways where the participants created a second layer for their badge (See Figure \ref{fig:teaser}). 

Even though the activity was only 90 minutes, the desire for iteration was evident. P24 and P63 expressed the need for flexibility and the opportunity to refine designs through iteration. This might be because some participants discovered that certain questions were easier to quantify and represent visually than others.

In conclusion, the Data Badge workshops not only provided participants with hands-on experience in data visualization but also offered insights into the creative, collaborative, and iterative nature of design processes. The balance between aesthetics and functionality, the dynamics of teamwork, and the creative use of limited resources emerged as central themes, highlighting the complexity and richness of engaging with data visualization.

\section{Discussion}

\subsection{Fostering Data-Driven Collaboration for Diverse Audiences}

The concept of data visualization often presents itself as a rigid and complex subject, with many perceiving a significant barrier to entry \cite{d2023data}. Co-badge offered an engaging and fun approach to overcome this obstacle while fostering data-driven collaboration. Our approach allowed participants to go through different stages of data visualization collaboratively, without a formal introduction, through learning by doing. Participants discovered critical steps and challenges in the data visualization process such as appropriately matching data types with visual variables or how subjective decisions in the process can profoundly impact the final outcome. 

Previous examples of data badges typically required data collection \cite{nissen2015data} or the preparation of custom materials for workshops \cite{panagiotidou2020data}. In contrast, the materials and processes for Co-Badge did not require any data collection or preparatory work beyond the purchase of stationery materials, thereby simplifying replication. Walny et al., (2015)\cite{walny2015exploratory} argue about the challenges participants face when attempting data visualization sketching with paper and pencils, particularly in exploring various iterations or fully representing the dataset. In contrast, our workshop, which incorporates the use of certain stationery materials, facilitates a smoother initiation into the creative process. This added to the playfulness of the workshop and created opportunities for learning as it lifted the obstacle of having certain skills such as sketching or getting familiar with new materials. For instance, the use of dot stickers in various sizes and colours emerged as a favoured option among participants, further easing the process. Although the short time frame for creating prototypes can introduce tension, emphasizing the process rather than the outcome and introducing materials that people were not afraid to disrupt, helped alleviate this pressure and enabled participants to focus on creating data visualisations and teamwork. Thus, presenting materials and processes that do not require any particular experience, even in the context of data visualization, may lift the barrier to active participation. 

Also, in our work, we had three diverse groups of participants of different personal and professional backgrounds. In almost all cases, insights related to data visualization principles emerged. As Bach et al. \cite{bach2023challenges} argue the future of data visualisation education needs to go beyond the classroom and involve diverse actors since data engagements are becoming more relevant for everyone in everyday life \cite{haddadi2016human}. Thus, it is important to design for a more inclusive space where we recognize data visualisation as a diverse and dynamic discipline allowing people to collaborate, learn, design and introduce their own perspectives. Incorporating a sense of play and using materials that participants are familiar with could foster an inviting and creative atmosphere.

Moving forward, we will explore using different materials and experiment with various collaboration structures, such as starting with individual tasks before group activities, to understand optimal approaches for learning data visualization. Additionally, targeting diverse groups, including children, could provide valuable insights into how different audiences engage with data visualization. Addressing these areas will refine Co-badge and expand its applicability, fostering a more inclusive and engaging environment for collaborative data visualization.

\subsection{Limitations}

Through the Co-Badge activity, we illustrated the intricacies of data-driven collaboration to a diverse audience, spanning novice to experienced in data visualization. However, participants from the same field faced challenges during the convergent parts of the process, such as selecting visual variables. To counter this, we recommend forming interdisciplinary groups to enhance the collaborative experience. Nonetheless, it is important to acknowledge that the short and fun nature of the exercise may not fully represent the dynamics of longer-term collaborations.

\section{Conclusion}

Overall, Co-badge offered diverse groups a playful and collaborative way to learn about the data visualization process. This activity can be applied in various settings, such as kick-off meetings for visualization projects, seminars, conferences, workshops, or hackathons. Through three workshops, participants from various disciplines engaged in a collaborative activity. Co-badge served as a vehicle to explore this topic collaboratively, aiming to take visualization education outside the traditional classroom context. By fostering a playful and accessible environment, this work is part of a larger effort that seeks to create new learning opportunities for diverse audiences, ultimately enhancing the inclusivity of data visualization education.

\bibliographystyle{ACM-Reference-Format}
\bibliography{sample-base}


\end{document}